\documentclass[floatfix,aps,prd,showpacs,amsmath,nofootinbib,preprintnumbers,
twocolumn]
{revtex4}
\usepackage{graphicx}
\usepackage{colordvi}
    \def\be{\begin{equation}}
    \def\ee{\end{equation}}
    \def\ba{\begin{eqnarray}}
    \def\ea{\end{eqnarray}}
\begin{document}

\title{Note on nonstationarity and accretion by primordial black holes in Brans-Dicke theory}

\author{B. Nayak}
\altaffiliation{bibeka@iopb.res.in}
\affiliation{Department of Physics, Utkal University, Vanivihar,
Bhubaneswar 751004, India}

\author{L. P. Singh}
\altaffiliation{lambodar\_uu@yahoo.co.in}
\affiliation{Department of Physics, Utkal University, Vanivihar,
Bhubaneswar 751004, India}

\begin{abstract}
We consider the evolution of primordial black holes by including nonstationarity in the formation process and accretion of radiation in Brans-Dicke theory.
Specifically, we focus on how $\eta$, the fraction of the horizon mass 
the black hole comprises capturing nonstationarity, affects the lifetimes of these primordial black holes. Our calculation reveals that the primordial black hole dynamics is controlled by the product $f\eta$ where $f$ is the accretion efficiency. 
We also estimate the impact of $\eta$ through  $f\eta$ on
the primordial black holes' initial mass fraction constraint obtained
from the $\gamma$-ray background limit.
\end{abstract}
\maketitle


In our previous paper \cite{nsm}, we showed that in Brans-Dicke(BD) theory accretion significantly
prolongates the lifetime of Primordial Black Holes(PBHs) by considering the initial mass of PBH as
horizon mass. This pertains to assuming a stationary flow of radiation onto the PBH. However, the real situation could be much more complicated \cite{imp} involving nonstationary processes. The numerical solution of nonstationary spherically symmetric problem of PBH formation \cite{imp, nn} suggests that PBH will have mass considerably smaller than the mass within the cosmological horizon at formation epoch. Also it has been shown \cite{imp, nn} that the 
nonstationarity of the PBH formation dynamics can indeed be captured by 
introduction of a parameter $\eta$ which measures the fraction of horizon 
mass aggregating to form the PBH. In our present work, we take $\eta$ as a free parameter and study its impact on PBH evolution dynamics. We also estimate the impact of $\eta$ in modifying the constraint on PBHs initial mass fraction
in BD theory obtained from the $\gamma$-ray background limit for
 PBHs evaporating now.

Using the calculation of our previous paper \cite{nsm}, one can get that the mass evolution of the PBH in the radiation-dominated era increases due to accretion as
\be \label{acc}
M(t)=M_i\Big[1 +\frac{3}{2}f\eta\Big(\frac{t_i}{t}-1\Big)\Big]^{-1}
\ee

where $M_i$ is the PBH mass at its formation time $t_i$ =$\eta M_H(t_i)$ with $M_H$ is the horizon mass, $f$ is the accretion efficiency and $\eta$ denotes the fraction of the horizon mass aggregating to form PBH.

For large times, the accreting mass approaches  its maximum value as
\be \label{mm}
M_{max}=\frac{M_i}{1-\frac{3}{2}f\eta}
\ee
 which gives that $f\eta < \frac{2}{3}$ . Thus, it is the product of $f$ and $\eta$ and not their individual values which controls PBH dynamics. This bound, obviously, is not as restrictive on $f$ as it was in our previous work \cite{nsm}. Here, in principle, $f$ can take a value equal to 1 while respecting the bound $f\eta< \frac{2}{3}$ .

The variation of accreting mass with $\eta$ is shown in the Figure-1. It is seen that mass of the PBH increases with $\eta$. However, for $\eta \leq 0.01$ the growth of PBH due to accretion is negligible.

\vskip 0.1in

\begin{figure}[h]
\includegraphics[scale=0.3]{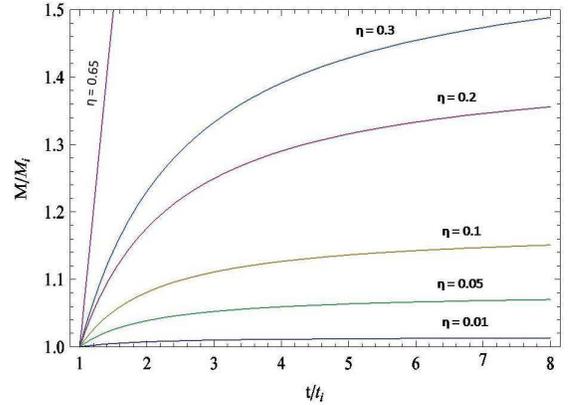}
\caption{Variation of accreting mass with different value of $\eta=0.65, 0.3, 0.2, 0.1, 0.05, 0.01$, but
with same accretion efficiency $f=1$.}
\label{fig1}
\end{figure}

Again, repeating the analysis of \cite{nsm}, we can write the evaporation time for those PBHs which are completely evaporated in radiation-dominated era as
\ba \label{rde}
t_{evap}=\Big(\frac{M_i}{1-\frac{3}{2}f \eta} \Big)^2 \Big[\frac{3}{2}fG_0\alpha^{-1}\Big(\frac{t_0}{t_e}\Big)^{3n} \nonumber \\ +\Big(3\alpha\Big)^{-1}\Big(\frac{t_0}{t_e}\Big)^{2n}\Big(\frac{M_i}{1-\frac{3}{2}f\eta}\Big)\Big]
\ea
where $\alpha=\frac{\sigma}{256 \pi^3 G_0^2}$ with $\sigma$ is the 
black body constant, $t_e \sim$ is the era of matter-radiation equality, 
$t_0 \sim$ is the present 
time, $G_0 \sim$ is the present value of $G$ $\simeq \frac{t_{pl}}{M_{pl}}$,
and $n$ is a parameter related to BD parameter $\omega$ as $n=\frac{2}{4+3\omega}$.
Since solar system observations \cite{bit} require that $\omega$ be large ($\omega \geq 10^4$), $n$ is very small ($n \leq 0.000$ $07$) .

The variation of evaporation time of these PBHs with $\eta$ as shown in the Figure-2 indicates that PBH live longer with an increase in the $\eta$ value. Here we have used $t_e=10^{11}s$ and $t_0=4.42 \times 10^{17}s$.

\vskip 0.1in

\begin{figure}[h]
\includegraphics[scale=0.3]{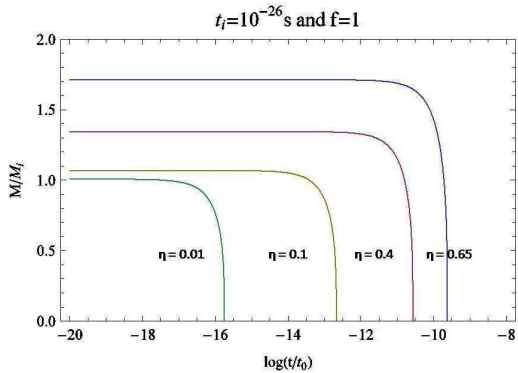}
\caption{Variation of evaporation time of PBH with formation time $10^{-26}sec$ for different $\eta=0.65, 0.4, 0.1, 0.01$, but
with same accretion efficiency $f=1$.}
\label{fig2}
\end{figure}

As shown in \cite{nsm}, we can express the  formation time for a PBH evaporating in matter-dominated era in terms of its evaporation time $t_{evap}$ as
\ba \label{mde}
t_i \approx G_0 \frac{1}{\eta} \Big\{1-\frac{3}{2}f \eta \Big\}\Big(\frac{t_{0}}{t_e}\Big)^n \times \Big[3\alpha\Big(\frac{t_e}{t_{0}}\Big)^{2n}t_e \nonumber \\\Big\{1+\Big(2n+1\Big)^{-1}\Big(\frac{t_{evap}}{t_e}\Big)^{2n+1}\Big\}\Big]^{\frac{1}{3}}  .\label{ft}
\ea
This equation also enables us to calculate the initial mass of evaporating PBH using the formula $M_i=\eta G^{-1}t_i$.

We now compute the formation times and the initial masses of the PBHs that
are evaporating at the present era and present the results in Table-I.
We have included $\eta=1$ case for $f=0$ for comparison with our previous result of Ref\cite{nsm}. It may be noted that for $\eta=1$, $f$ cannot exceed the value of $2/3$.

\begin{table}
\begin{tabular}[c]{|c|c|c|c|c|}
\hline
\multicolumn{5}{|c|}{$f=0$ \hskip 0.6in $f=1$}\\
\hline
$\eta$  &  $t_i \times 10^{-23}$s &  $M_i \times 10^{15}$g  &  $t_i\times 10^{-23}$s  &  $M_i\times 10^{15}$g\\
\hline
$1$  &  $2.369$  &  $2.366$  &  & \\
\hline 
$3/5$  &  $3.949$   &  $2.366$ & $0.395$ & $0.237$\\
\hline
$1/2$  &  $4.739$  &  $2.366$ & $1.185$ & $0.592$\\
\hline
$1/5$  &  $11.847$  &  $2.366$ & $8.293$ & $1.657$\\
\hline
$1/10$  &  $23.693$  &  $2.366$ & $20.148$ & $2.013$\\
\hline
$1/100$   &  $236.934$  &  $2.366$ & $233.384$ & $2.331$\\
\hline
\end{tabular}
\caption{The formation times and initial masses of PBHs that are evaporating today corresponding to
$f=0$ and $f=1$ for different values of $\eta$.}
\end{table}

We also estimate the constraint on the PBH mass fraction for presently evaporating PBHs which gives $\gamma$-ray background, as follows

The fraction of the Universe going into PBHs with formation mass $M_i$ is \cite{nsm}
\be \label{bet4}
\beta(M_i) < 10^{-4} \times \Big(\frac{M_i}{M_e}\Big)^{\frac{1}{2}} \times \Big(\frac{t_e}{t_0}\Big)^{\frac{2-n}{3}}  
\ee
where $M_e$ is the initial mass of PBH formed at time $t_e$.\\
Using the values of $t_e$, $t_0$ and $n$, we get
\be \label{bet5}
\beta(M_i)<1.174 \times 10^{-33} \times \eta^{-\frac{1}{2}} \times \Big(\frac{M_i}{g}\Big)^{\frac{1}{2}}
\ee

We can now use the expressions for $M_i(\eta G^{-1}t_i)$  in terms of the evaporation
time $t_{evap} = t_0$ to obtain the values of $\beta(M_i)$ corresponding to different  values of $\eta$. These are displayed in
Table~II. Here again we have included $\eta=1$ case for comparison with our previous result \cite{nsm}. From Table-II, we observe that increase in value of $\eta$ makes the
limits on the initial mass fraction more stringent. It may be noted that
similar considerations would also apply to constraints on the initial
mass fraction $\beta(M_i)$ obtained from other physical considerations
such as those due to entropy or nucleosynthesis bounds.

\begin{table}
\begin{tabular}[c]{|c|c|c|c|c|}
\hline
\multicolumn{5}{|c|}{$f=0$ \hskip 0.6in $f=1$}\\
\hline
$\eta$  &  $M_i \times 10^{15}$g  &  $\beta(M_i) <$ & $M_i \times 10^{15}$g  &  $\beta(M_i) <$ \\
\hline
$1$  &  $2.366$ & $5.71 \times 10^{-26}$ &  & \\
\hline
$3/5$  &  $2.366$ & $7.371 \times 10^{-26}$ &  $0.237$   &  $2.333 \times 10^{-26}$\\
\hline
$1/2$  & $2.366$ & $8.075 \times 10^{-26}$  &  $0.592$  &  $4.040 \times 10^{-26}$\\
\hline
$1/5$  & $2.366$ & $12.768 \times 10^{-26}$  &  $1.657$  &  $10.687 \times 10^{-26}$\\
\hline
$1/10$  & $2.366$ & $18.056 \times 10^{-26}$ &  $2.013$  &  $16.658 \times 10^{-26}$\\
\hline
$1/100$  &  $2.366$ & $57.1 \times 10^{-26}$  &  $2.331$  &  $56.687 \times 10^{-26}$\\
\hline
\end{tabular}
\caption{Upper bounds on the initial mass fraction of PBHs that are
evaporating today for different values of $\eta$ at accretion efficiency $f=1$.}
\end{table}

In this paper we have considered the evolution of primordial black holes in Brans-Dicke theory by parametrizing the nonstationarity in the formation mechanism in terms of a parameter $\eta$ and including accretion of radiation.
As an extension of our previous work \cite{nsm}, here we study the effects of initial
mass of PBH and accretion efficiency on their lifetimes rather than their dependence on accretion efficiency alone.
Our calculation shows that the accretion efficiency $f$ and the fraction  of horizon mass PBH comprising ($\eta$) are inter-related as $\eta f < 2/3$. So for accretion to be effective, the product of $f$ and $\eta$ should be nearly $2/3$. Since $f$ can not exceed 1, for PBHs significantly smaller than horizon, accretion is negligible. For large value of $\eta$, the lifetimes of PBHs prolongate due to significant accretion.
This can be explained by the fact that a large PBH will tend to capture a lot of neighbouring material through its' gravitational effect compared with smaller ones.

The cosmological evolution of PBHs could lead to various interesting
consequences during different eras. Here we use the observational
limits on the $\gamma$-ray background \cite{mac and carr} to compute the effect of $\eta$ on constraining the initial mass fraction of the PBHs.
We found that the incrase in the value of $\eta$ makes the limits on the initial PBH mass fraction more stringent.

Thus, the effectiveness of accretion is controlled by the product $f\eta$ and hence the lifetimes for PBHs could be enhanced depending upon the value of $\eta$, which in turn modifies the constraints on initial mass fraction of PBHs.\\

B. Nayak is thankful to the Council of Scientific and Industrial Research, Government of India, for SRF, F.No. $09/173(0125)/2007-EMR-I$ .


\end{document}